\documentclass[sigconf,nonacm]{acmart}

\usepackage[skip=1pt]{caption}
\usepackage{algorithm}
\usepackage{algpseudocode}
\usepackage{multirow}
\usepackage{color}
\usepackage{tikz}
\newcommand*\circled[1]{\tikz[baseline=(char.base)]{
            \node[shape=circle,draw,inner sep=1pt] (char) {#1};}}
\definecolor{Blue} {rgb} {0.3,0.3,0.9}

\newcommand{\NDH}{\textsc{HOST} }
\newcommand{\NDR}{\textsc{SHARED} }
\newcommand{\NDW}{\textsc{UNIQUE} }
\newcommand{\NDD}{\textsc{REP} }
\newcommand{\UOP}{$\mu$op }
\newcommand{\UOPS}{$\mu$ops }

\begin{document}

\title{NDS: Programmer-Free Offload of High-Performance Near Data Strands}
\titlenote{This manuscript is an extended version of the two-page PACT '26 poster paper (DOI: 10.1145/3838684.3846876).}

\author{Shreyas Singh}
\affiliation{\institution{University of Utah}\country{USA}}
\author{Pratyush Nandi}
\affiliation{\institution{University of Utah}\country{USA}}
\author{Lin Jia}
\affiliation{\institution{University of Utah}\country{USA}}
\author{Shankar Balachandran}
\affiliation{\institution{Intel}\country{India}}
\author{Rajeev Balasubramonian}
\affiliation{\institution{University of Utah}\country{USA}}

\begin{abstract}
Near Data Processing (NDP) has the potential to significantly improve system performance and energy by alleviating data movement bottlenecks. However, most NDP proposals pose heavy requirements for the software stack, data layout, and/or the underlying hardware. To broaden NDP adoption, this work focuses on a modular hardware-centric approach that automates the above steps and can target unmodified binaries.

This paper presents Near Data Strands (NDS), a framework that orchestrates tasks and data automatically without involving the programmer and the software stack. As a first step towards this ambitious goal, this work focuses on regular loops that meet specific criteria. The framework identifies potential offloadable loops, decomposes them into parallel execution strands, creates a high performance instruction schedule, performs the required data marshaling and orchestration, and initiates near-data execution. We demonstrate that NDS achieves a 3.14$\times$ geomean speedup over a single-core host baseline and a 1.82$\times$ geomean speedup over 8-core host-parallel variants, all without any modifications to the software stack. These gains grow as repeated loop invocations amortize offload costs, showing that legacy applications can leverage a transparent hardware approach to extract benefits from NDP.
\end{abstract}

\maketitle

\section{Introduction}
\label{intro}


Near Data Processing (NDP) has the potential to improve system performance and energy by reducing data movement, overcoming pin bandwidth limitations, and exposing high memory-side parallelism~\cite{shafiee2016isaac, huangfu2022beacon}. NDP is also becoming increasingly concrete, with prototypes and products from Samsung~\cite{kim2021aquabolt,ke2021nearaxdimm}, SK Hynix~\cite{he2020newton}, and UPMEM~\cite{devaux2019upmem}. Recent interest has grown further because many AI kernels, including GEMV-heavy phases in LLM inference, are constrained by memory bandwidth; both industry~\cite{park2024lpddr,ortega2024pim} and academia~\cite{yun2024duplex,choi2023unleashing,park2024attacc,li2024specpim,heo2024neupims,gu2025pim} have therefore proposed NDP accelerators for these workloads.



The broader question is whether this near-memory capability can be used beyond carefully hand-tuned accelerators. Many business- and science-critical applications contain memory-bound kernels, including phases in genomic analysis~\cite{chen2023uppipe,diab2023framework,kim2018grim}, database scans and simple queries over large datasets~\cite{bernhardt2023pimdb,lim2023design}, and cryptographic workloads with large working sets~\cite{dong2024toleo,nejatollahi2020cryptopim}. However, converting these applications to use NDP is difficult because current systems often require programmer-visible data movement, new offload APIs, compiler support, ISA changes, or NDP-specific data layouts.

This paper studies a concrete question: can hardware transparently identify useful regular loops from unmodified binaries, then offload them when loop structure, data placement, and reuse behavior make near-data execution profitable? The target is a constrained but common class of repeated, regular loops that can be detected, transformed into parallel strands, mapped to near-data cores, and executed profitably while unsupported or low-benefit loops continue on the host.

Cost visibility is central to this framing. Transparent offload is useful when the benefits of memory-side bandwidth and parallelism exceed the costs of detection, template storage, page movement, coherence actions, MMU/OS handoff, and fallback. This work defines the hardware mechanisms and conditions under which those costs can be amortized.



We introduce Near Data Strands (NDS), an architecture in which conventional binaries first execute on a host processor and hardware identifies repeated regular loops that are candidates for NDP. In line with current commercial directions~\cite{devaux2019upmem}, we assume simple in-order near-data cores (NDCores) with small area and energy footprints are placed near DRAM banks. As the binary runs, a host-side hardware engine observes loop execution, records a template for loops that satisfy structural and memory-access checks, and later instantiates parallel strands from that template for execution across NDCores. Because the host has already executed the loop on an out-of-order core, the hardware can observe runtime information about iteration counts, touched addresses, dependence behavior, and micro-ops
that extend beyond static compiler visibility. NDS uses this information to generate parallel strands, construct a higher-ILP micro-op schedule for in-order NDCores, and guide page-aware data marshaling. The framework falls back to host execution when a loop diverges from the recorded template or fails safety checks.


To develop this scoped NDS architecture, this paper addresses four sub-problems: (i) detecting candidate loops in hardware, (ii) creating parallel strands from accepted loop templates, (iii) orchestrating page movement and coherence at offload boundaries, and (iv) improving in-order NDP execution using schedules learned from the out-of-order host. The key opportunity is to combine memory-side bandwidth and bank-level parallelism with enough hardware introspection to make simple NDCores effective on repeated regular loops.

This paper makes the following contributions:
\begin{itemize}
\item A hardware-centric NDP architecture that preserves the traditional programming model while targeting repeated, regular loops in unmodified program binaries.
\item Techniques for identifying candidate loops, checking offload eligibility, and automatically creating parallel strands that can be offloaded to NDCores.
\item An introspective scheduling technique that observes out-of-order host execution to create higher-ILP micro-op schedules for simple in-order NDCores.
\item Page-aware data marshaling and coherence mechanisms that make offload costs, data affinity, and fallback conditions explicit.
\item An evaluation across 12 kernels showing that NDS detects all dynamically observed loops, identifies 12 inner-loop and 3 outer-loop offload candidates, and improves performance by 3.14$\times$ geomean over a host baseline and 1.82$\times$ geomean over 8-core host-parallel variants.
\end{itemize}

\section{Motivation and Background}
\label{sec:motivation}

\subsection{NDP Approaches and Hardware}
\label{sec:motndp}

Researchers have explored near-data processing for decades. Across these proposals, four design choices are especially relevant to NDS: how computation is performed, where it is placed, what granularity is offloaded, and how general the near-data compute unit is.


\subsubsection*{Mode of Computation}

Based on how computation happens, we categorize NDP approaches as processing using memory, processing in memory, and processing near memory. Processing-\emph{Using}-Memory exploits memory cell and bitline properties to perform computations within arrays, sometimes with new sense-amp circuits~\cite{seshadri2017ambit,wang201928computesram,shafiee2016isaac,gao2019computedram}. While such computations are often bit-serial, high performance is achieved with very high bitline-level parallelism.


Processing-\emph{In}-Memory proposals place compute units adjacent to memory arrays, benefiting from bank-level parallelism and bandwidth~\cite{devaux2019upmem, kwon2023skhynix, kim2021aquabolt}. Given DRAM constraints, these compute units are often simple. Processing-\emph{Near}-Memory proposals can use larger compute units outside the memory die but closer than the host CPU, providing more bandwidth than host-side execution observes~\cite{ke2021nearaxdimm,huangfu2019medal}.

\subsubsection*{Locations of Computations}
Computation can be placed at multiple locations throughout the memory hierarchy: near, in, or using caches~\cite{wang201928computesram,nori2021reduct,gauchi2020reconfigurable}; near, in, or using memory~\cite{devaux2019upmem,kwon202125,lee20221ynm,ke2021nearaxdimm}; and even near storage~\cite{wilkening2021recssd, wang2024beacongnn,zhang2024omnicache,seneviratne2023nearpm}.
Some proposals provide NDP at multiple locations in the hierarchy~\cite{lockerman2020livia,schwedock2022tako,nori2021reduct}. The location determines both the available bandwidth and the software/hardware contract required to move work and data.


\subsubsection*{Computation Offload Granularity}
NDP proposals can also be categorized by offload granularity. Fine-grained proposals~\cite{ahn2015pei,lockerman2020livia,kim2021aquabolt} involve the host coordinating a single instruction or a small group of instructions near memory. Coarse-grained proposals offload larger regions of code to near-data units~\cite{devaux2019upmem,devic2022topim}. This choice determines whether complexity appears in the host, the near-data unit, the compiler, or the programmer-visible offload interface.



\subsubsection*{Generality of Compute}
NDP proposals also span a spectrum of compute generality. Highly specialized architectures, such as SK Hynix's Newton~\cite{kwon2023skhynix,lee2019design,lee20221ynm}, can offer high performance for specific workloads by specializing tightly to those kernels. General-purpose systems such as UPMEM expose in-order cores that can run broader code~\cite{devaux2019upmem}, while still requiring explicit programmer orchestration and data movement. This creates an adoption gap: flexible NDP hardware exists, but many deployed binaries and libraries face high integration barriers.




\subsubsection*{Scope of This Work}
This paper focuses on a Processing-\emph{In}-Memory architecture where each DRAM bank is associated with a simple in-order core, called an NDCore, and the host offloads entire loop regions to those cores. We use this target because it exposes the central tradeoff for transparent NDP: bank-level parallelism and memory-side bandwidth are attractive, but orchestration, page placement, coherence, and setup costs must be controlled. The mechanisms are not tied to one commercial system, but this model gives the paper a concrete hardware target and cost model.

\subsection{Loops as NDP-Offload Candidates}
\label{sec:motloop}

The goal of NDS is to identify NDP-offloadable code at runtime from an unmodified binary. Regular loops are a deliberate first target because they combine three properties that transparent offload needs: predictable control flow, analyzable memory behavior, and repeated execution that can amortize setup costs.
Loops are also a natural hardware target. Prior work has shown that a significant fraction of dynamic execution occurs in loops, even though loops represent a small fraction of static instructions~\cite{padmanabha2015dynamos, padmanabha2017mirage, mcfarlin2013discerning, palomar2009reusing}. Modern processors already include hardware such as Loop Stream Detectors~\cite{intellsd} to identify loops for energy efficiency, suggesting that lightweight loop recognition is a reasonable starting point. NDS extends this idea from recognizing loops to determining whether a loop has simple induction behavior, supported operations, regular memory accesses, enough parallelism, and a favorable cost profile for NDP.
This scope is intentionally limited to loops with predictable control and regular memory behavior. Pointer chasing, highly irregular graph traversal, non-constant access strides, deeply data-dependent control flow, and short loops with little reuse are better matched to programmer, compiler, runtime, or data-layout support than to the transparent hardware mechanism studied here.


\subsection{Hardware-Centric Offloading Scheme}
\label{sec:mothardware}

NDS adopts a hardware-centric approach to reach deployed binaries, closed-source libraries, and legacy applications without requiring source changes. A hardware mechanism can observe the actual dynamic execution path, the micro-ops produced by the host frontend, the addresses generated by the running binary, and the behavior of repeated loop invocations. This runtime view enables cost/benefit decisions similar in spirit to prior hardware introspection mechanisms~\cite{vijaykumar2018case}, while also exposing micro-op-level scheduling opportunities that are unavailable to compilers and source-level tools~\cite{tran2018swoop}.
The transparent hardware stance also forces a strict contract: NDS accepts loops whose behavior it can validate cheaply, maps data before offload begins, and falls back to normal host execution when those conditions are not met. This contract keeps the proposal scoped as a feasibility study with explicit eligibility and fallback conditions.

\section{The NDS Hardware Framework}
\label{sec:proposal}

\subsection{Overview}
\label{sec:overview}

\begin{figure}
    \centering
    \includegraphics[width=0.9\linewidth]{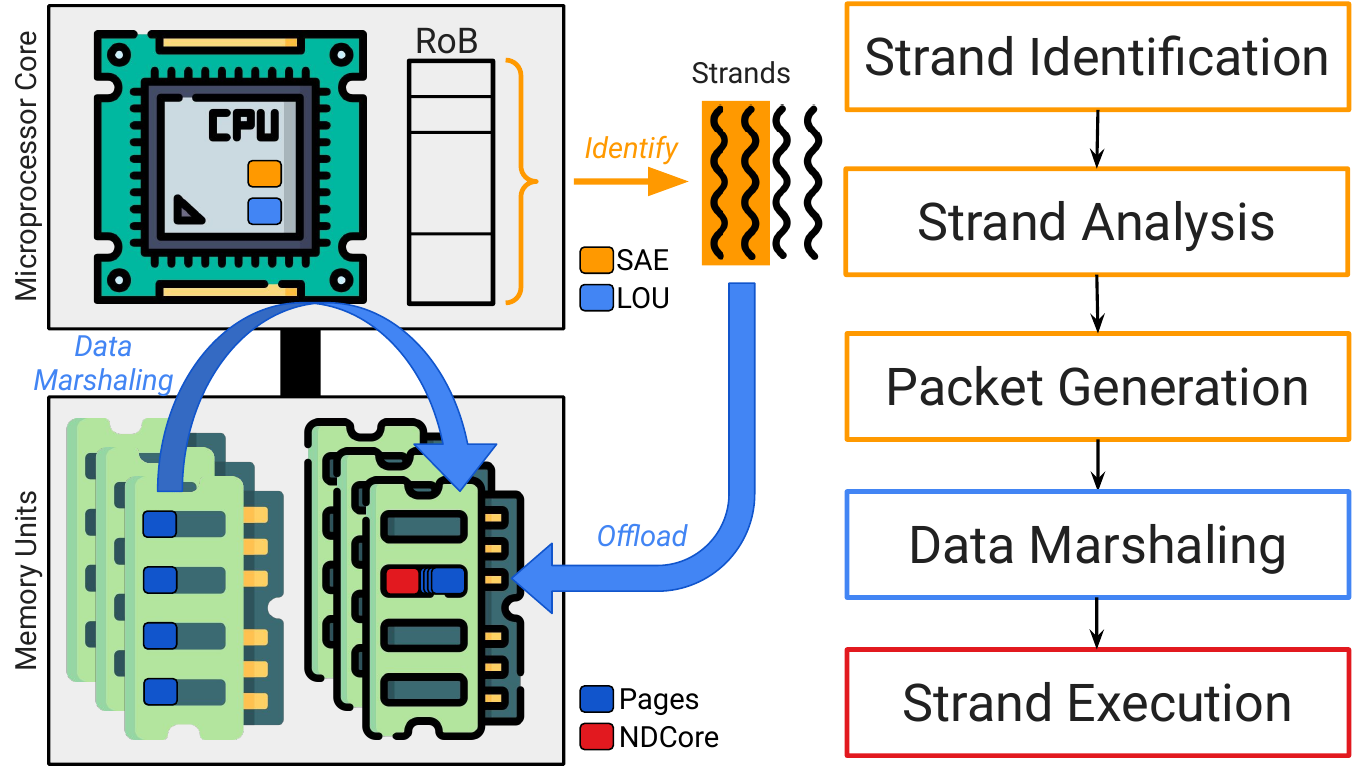}
    \caption{Steps involved in the NDS Framework}
    \Description{Pipeline overview showing loop detection, analysis, template generation, data marshaling, and near-data execution.}
    \label{fig:overview}
\end{figure}

In the proposed NDS architecture, a hardware unit in the host processor analyzes the executing program and identifies regular loops that are good candidates for NDP execution. During an initial invocation, the hardware records the loop body and associated metadata as a \textit{template}. On a later invocation, the template is used to create multiple \textit{strands}, map their data to the appropriate NDCores, and execute them in parallel. If the loop no longer matches the recorded template, or if its data cannot be mapped safely, the loop executes on the host.

We organize the NDS framework into five key components, each discussed in the next sub-sections: (i) strand detection, (ii) strand analysis, (iii) parallelization and template generation, (iv) data marshaling, and (v) strand execution.

We focus exclusively on simple loops. We rely on a hardware engine alongside every host out-of-order core that observes the dynamic instruction stream and records metadata about candidate loops. We refer to this unit as the Strand Analysis Engine (SAE, see Figure~\ref{fig:overview}). The SAE identifies loops likely to benefit from NDP: loops with deterministic iteration spaces, supported operations, regular memory accesses, high data affinity, and enough expected parallelism to amortize offload costs. The SAE then generates a \textit{template} that represents the loop and is used for future offloads.

When a future invocation matches an existing template, host execution pauses at the loop boundary while the Loop Orchestration Unit (LOU) performs the required orchestration. The LOU assigns iterations and their pages to specific NDCores, initializes each strand with the correct live-in values and iteration bounds, and ensures that each strand can execute using local or replicated pages. Once all NDCores signal completion, the host resumes execution.
Unlike thread-level speculation~\cite{steffan2000scalable,estebanez2016survey}, strands launched by NDS are not speculative; all dependence and data-placement checks are completed before near-data execution begins.


Figure~\ref{fig:detect_fsm} illustrates the progression of a loop candidate from initial detection \circled{1} \circled{2} and profiling \circled{3} \circled{4} to the final verification \circled{5} and data marshaling performed by the LOU. Table~\ref{tab:checks} provides a summary of the introduced hardware components and the checks they must perform. We will refer to these checks as we next describe each component in detail.

\subsection{Strand Detection}
\label{sec:strdetect}

\begin{figure}
  \begin{minipage}{0.49\linewidth}
    \includegraphics[width=\linewidth]{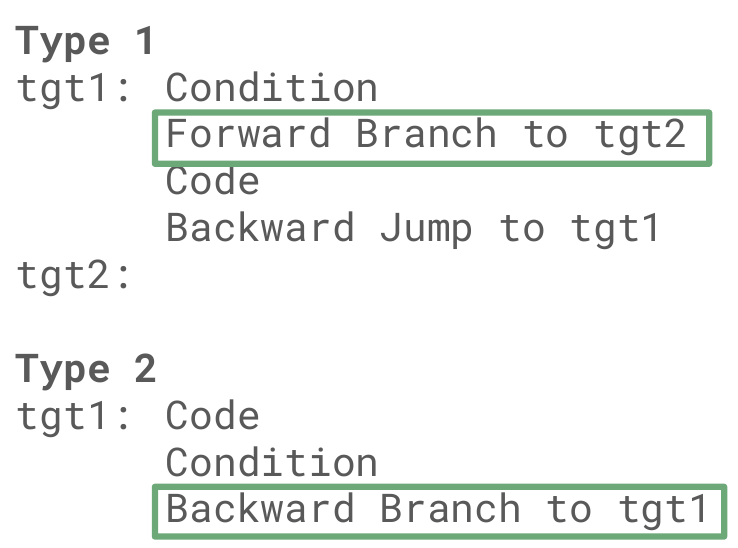}
  \end{minipage}\hfill
  \begin{minipage}[c]{0.5\linewidth}
    \caption{Two types of loops detected by NDS. A forward branch associated with a backward jump and a backward branch.} 
    \Description{Two control-flow patterns for loop detection: a forward branch followed by a backward jump, and a backward branch.}
    \label{fig:loop_examples}
  \end{minipage}
\end{figure}

\begin{figure}
    \centering
    \includegraphics[width=0.9\linewidth]{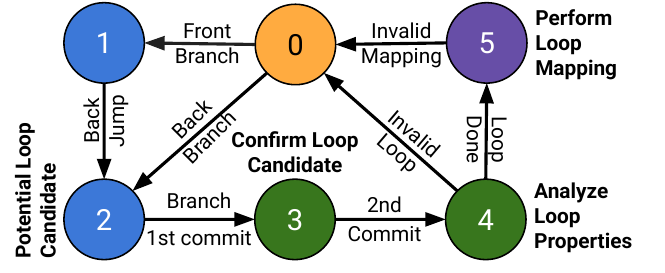}
    \caption{State-flow diagram of the SAE's offload process.}
    \Description{Finite-state machine for identifying, profiling, verifying, and offloading loop candidates.}
    \label{fig:detect_fsm}
\end{figure}

We first discuss identification of strands. As mentioned in Section \ref{sec:motloop}, many modern processors have dedicated Loop Stream Detectors that intercept executing loops to save power and energy. Our proposal builds on such existing hardware and augments it to track additional metadata and exploit NDP.  

\begin{figure}
    \centering
    \includegraphics[width=\linewidth]{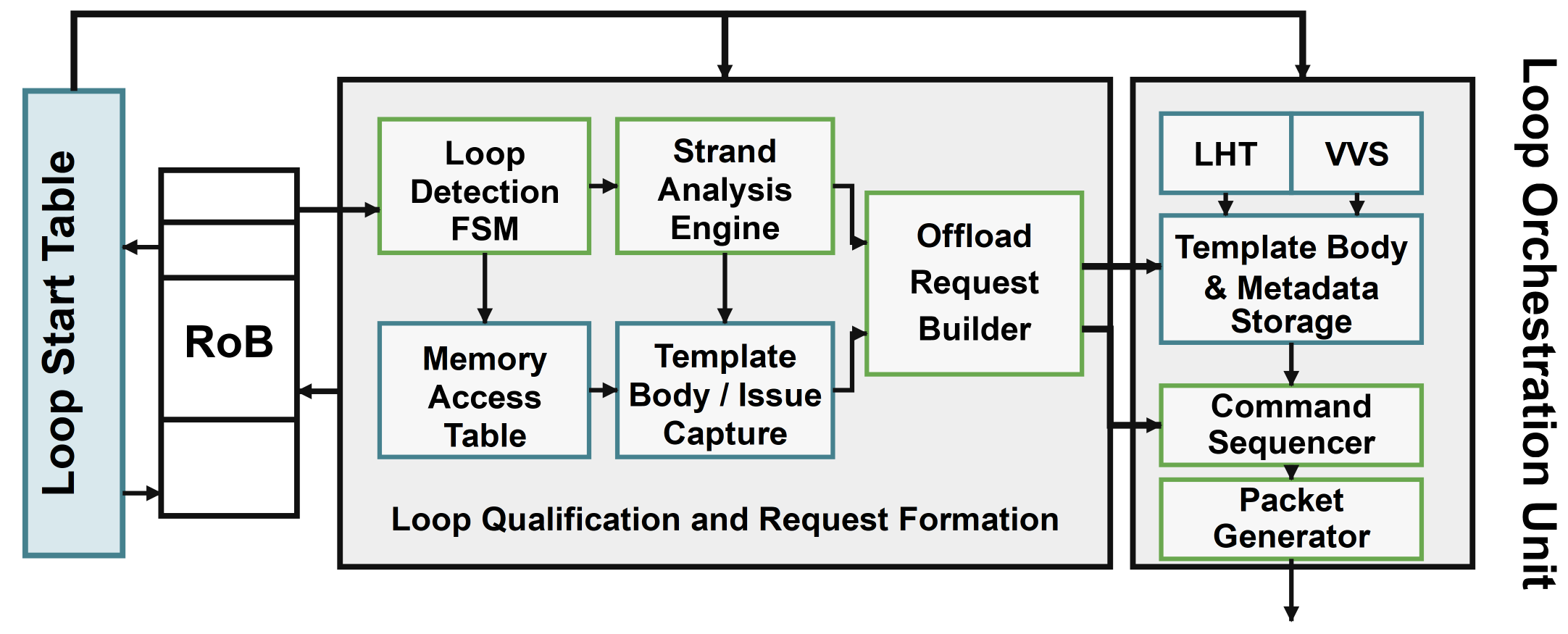}
    \caption{Structures needed on the Host CPU to enable the NDStrands Framework}
    \Description{Host CPU structures for loop detection, strand analysis, template storage, and loop orchestration.}
    \label{fig:host_hardware}
\end{figure}

Our loop detection mechanism (part of the Strand Analysis Engine in Figure \ref{fig:host_hardware}) is a Finite State Machine (FSM) that monitors \UOPS at Dispatch to identify two loop patterns (Fig \ref{fig:loop_examples}). The FSM's state transitions are triggered by branch \UOPS: (i) A forward branch (Type 1) transitions the FSM to an intermediate state \circled{1}, where it awaits a subsequent unconditional backward jump to complete the pattern \circled{2}. (ii) A backward branch (Type 2) transitions the FSM directly to the "potential candidate" state \circled{2}.
In both patterns, the target of the final backward branch/jump is recorded as the \textit{Loop Start IP}. Once the branch commits and signals the end of the first iteration, the FSM promotes the loop to a loop candidate \circled{3} and signals the Loop Analysis Engine (LAE) to initiate profiling.


\begin{figure}
    \centering
    \includegraphics[width=\linewidth]{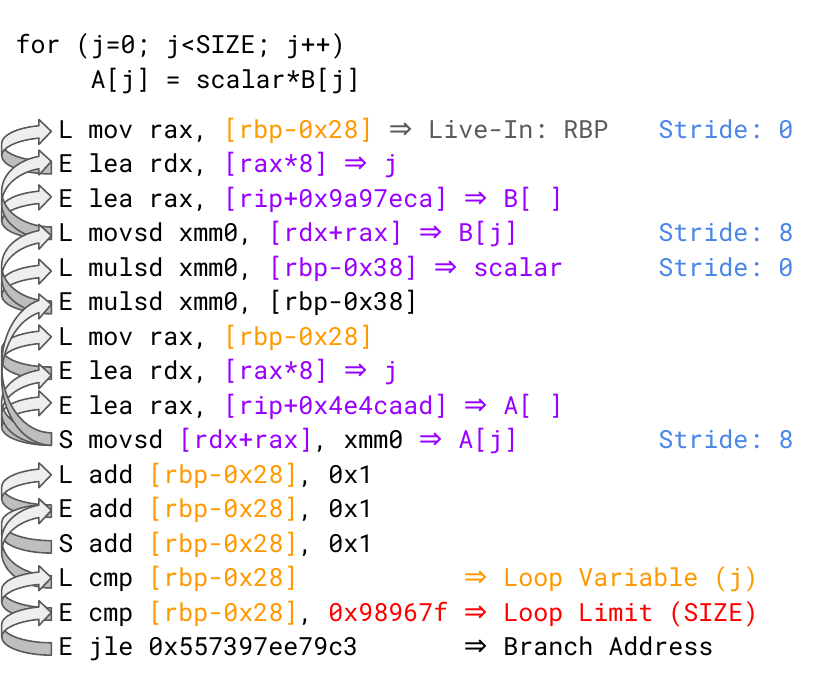}
    \caption{Code sequence and SAE analysis for the STREAM-Scale workload}
    \Description{Assembly-level STREAM-Scale loop annotated with the analysis information collected by the SAE.}
    \label{fig:code_stream}
\end{figure}

\begin{figure}
  \begin{minipage}{0.43\linewidth}
    \includegraphics[width=\linewidth]{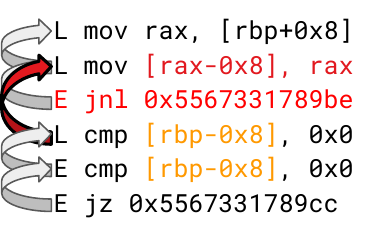}
  \end{minipage}\hfill
  \begin{minipage}[c]{0.55\linewidth}
    \caption{Code sequence where the SAE analysis fails as the Loop Variable update is non-deterministic and the loop body contains a branch.} 
    \Description{Assembly example of a rejected loop with non-deterministic loop-variable update and internal branch behavior.}
    \label{fig:code_fail}
  \end{minipage}
\end{figure}



\subsection{Strand Analysis}
\label{sec:stranalysis}

Once the detector identifies a candidate loop, the SAE profiles a bounded number of dynamic iterations to decide whether the loop can be represented as an NDS template. As \UOPS pass through the core, the SAE records compact metadata about the loop body, including opcode classes, register dependences, live-in values, issue order, load/store addresses, and loop-branch information. The loop analysis is performed after the profiled loop instance has executed, using the recorded metadata, and is therefore off the normal execution critical path. If the loop exceeds the supported template size or requires unsupported operations, the candidate is dropped and later executions proceed normally on the host.

During this profiling phase (\circled{3} \circled{4}), the Loop Analysis Engine (LAE) performs three checks. First, it validates resources: the loop body must fit in the template storage, the required live-in values and registers must fit within the NDCore limits, and all \UOPS must be supported by the NDCore. Second, it populates the Memory Access Table (MAT) and verifies that each memory operation has a constant stride across profiled iterations. Third, it checks that the loop exposes a deterministic iteration space.

The iteration-space check starts from the branch and compare \UOPS that close the loop and performs a bounded walk over the captured dependence links to identify the \textit{Loop Variable} (LV), the \textit{Loop Limit} (LL), and the \textit{Update} operation. The walk is limited by the profiled loop body and the configured template capacity. The SAE accepts the loop only when the LV update is constant, the LL is loop-invariant, and the values needed to reconstruct the induction variable are either constants or live-ins. These conditions allow the LOU to divide the iteration space into strands on later invocations.

Figure \ref{fig:code_stream} presents an example of a loop where the above analysis passes while Figure \ref{fig:code_fail} describes a case where it fails.



\subsection{Template Generation}
\label{sec:pktgenerate}

The NDS framework creates a template from loop candidates that successfully finish their profiling runs \circled{4}$\rightarrow$\circled{5}. The Loop Orchestration Unit (LOU) leverages these templates to verify and orchestrate NDP offloads on future invocations of the loop. NDS uses a hierarchical set of structures to manage these templates.

This process begins by creating an entry in the Loop Header Table (LHT), which stores the loop's invariant template. The template contains the parameterized loop body, the \UOP schedule (Section~\ref{sec:schedule}), the loop variable, loop limit, update operation, and the memory access descriptors collected in the MAT. Each memory descriptor records whether the access is a read or write, whether it is loop-invariant or strided, and the base and stride information needed to identify the pages touched by each iteration. The LHT also stores the list of live-in registers and memory values that form the \textit{version key} for this template.

The version-specific values are stored separately in the Version Value Store (VVS), pointed to by the LHT. The first VVS entry stores the live-in values, detected strides, and loop limit observed during the profiling run. Finally, the SAE adds the \textit{Loop Start IP} (Section~\ref{sec:strdetect}) to the Loop Start Table (LST), allowing later invocations of the same loop to find the template.

\begin{figure}
    \centering
    \includegraphics[width=\linewidth]{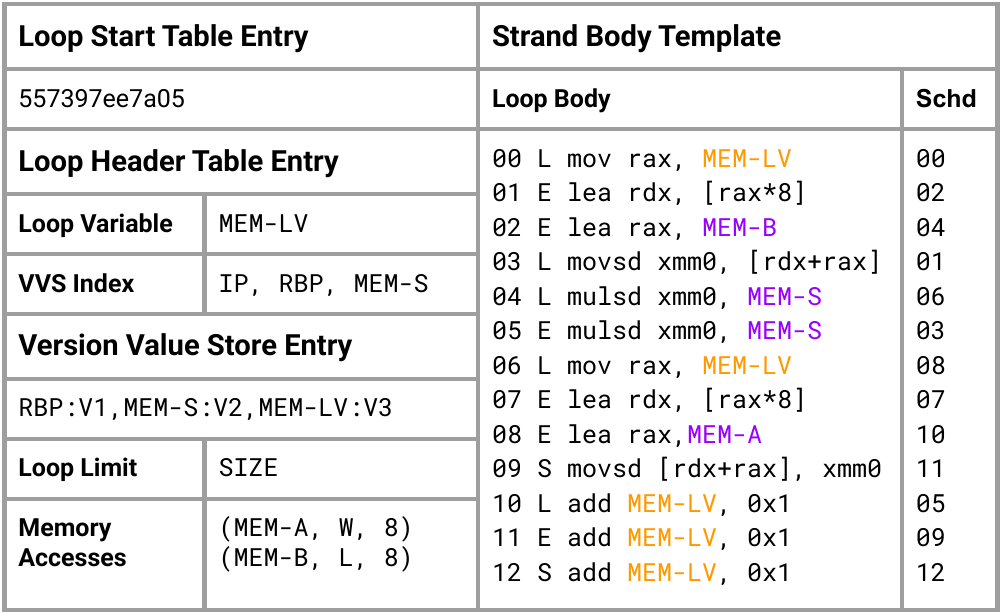}
    \caption{Entries related to a template generated for the STREAM-Scale Workload}
    \Description{Template entries for a STREAM-Scale loop, including loop metadata, live-in values, and memory access information.}
    \label{fig:packet_stream}
\end{figure}

The LST/LHT/VVS hierarchy (see the example in Figure~\ref{fig:packet_stream}, and additional details in Table~\ref{tab:checks}) is the core of NDS' loop version management. A hit in the LST at dispatch tags a possible loop start. When the corresponding branch commits, the LHT provides the version-key template, and the current architectural values are compared against the VVS. An exact match identifies the template and version to offload. A miss indicates that the current invocation is a different loop version; the SAE can profile it as a new version if resources are available, otherwise the loop executes on the host.

\subsection{Loop Orchestration and Data Marshaling}
\label{sec:looporchestration}

Near-data execution is constrained by bank locality: NDCores cannot access data outside their attached bank~\cite{rai2021design}, so each strand must be launched where its data is present. The Loop Orchestration Unit (LOU) maps iterations and pages to banks using the memory-access metadata in the strand template. It performs this orchestration at page granularity, which keeps metadata compact, aligns with OS allocation and migration mechanisms, and expresses data placement as physical-page constraints. Similar page-level control appears in cache coloring~\cite{zhang2009towards}, transparent and near-data page placement~\cite{almaruf2023tpp,wang2023affinity}, and fine-grained in-DRAM data relocation and caching~\cite{wang2020figaro}.

The LOU does not rewrite page tables directly. Instead, it computes the desired bank for each page and hands placement/remap intent to the privileged host MMU/OS path, which allocates or migrates pages to matching physical frames, updates virtual-to-physical mappings, and performs TLB invalidations. The NDCore launch packet then carries physical page/bank descriptors plus per-strand offsets. This mirrors tiered-memory and device-memory systems that extend OS page-management paths rather than bypassing them~\cite{lee2024pim,yan2019nimble,xiang2024nomad}. If the mapping cannot be committed safely, NDS aborts the offload before launch and executes the loop on the host.

The mapper is write-page centric. A page written by an offloaded strand has a single owning NDCore bank during an offload, while read-only pages may be replicated. Thus, write pages define ownership constraints, and read pages define replication or movement cost. NDS tracks four page states using 2 bits in the reserved physical-page-number bits~\cite{zhang2009towards}: \NDH, not managed by the LOU; \NDR, mapped to one NDCore for reads; \NDD, mapped to multiple NDCores for reads; and \NDW, mapped to one NDCore for writes. This encoding couples the data's bank, state, and physical address, making the placement visible to all hardware agents.

Algorithm~\ref{alg:looporchestration} summarizes the LOU mapping policy invoked at each offload. The LOU first estimates the number of store pages from the write-stream descriptors in the template and converts this into a store-page-per-bank quota. It then walks the iteration space, computes the all-pages, write-pages, and read-only pages touched by the current iteration, and chooses a target bank. If an iteration touches pages with existing write owners, the LOU chooses among the required owner banks using work balance and read-page replication cost as tie-breakers. If no existing owner constrains the iteration, the mapper either keeps the current region bank for a repeated write-page signature or advances to another bank when the store-page quota would be exceeded. Once a bank is selected, all touched pages are made present in that bank, and any newly written pages acquire that bank as their owner.

This policy handles loop-carried dependences that are expressible through page-level ownership. \textit{RAW} and \textit{WAW} dependences on pages written by offloaded strands are serialized by co-locating the dependent work at the owning bank, while independent write-page regions can still execute on different banks. Read-only pages may be replicated to reduce movement; if an analyzed loop would require an unsafe page-state transition or a dependence pattern that cannot be represented by this page-level model, the LOU discards the in-flight mapping and executes the loop on the host.

The mapper also records the page-to-bank placement, number of banks used, replicated read pages, and contiguous iteration steps assigned to each bank. These fields are used to estimate marshaling cost and NDCore launch work. The marshaling cost is heaviest on the first offload of the loop as pages are placed into the required states and banks; subsequent invocations can reuse this placement unless intervening host execution dirties replicated pages, as discussed next. Since the host MMU resolves translations before offload and the LOU uses physical page numbers to map strands to NDCores, the NDCores do not require a TLB.

\begin{algorithm}
\caption{LOU Mapping Policy}\label{alg:looporchestration}
\begin{algorithmic}
\State $SPB \gets \lceil \textproc{EstimateStorePages}(template)/N_{banks} \rceil$
\State $owner \gets \emptyset$ \Comment{write page $\rightarrow$ owning bank}
\State $current \gets 0$
\ForAll{iteration $i$}
    \State $Pages \gets \textproc{PagesTouched}(template,i)$
    \State $WritePages \gets \textproc{WrittenPages}(Pages)$
    \State $ReadPages \gets Pages - WritePages$
    \State $OwnerBanks \gets \{owner[p] \mid p \in Pages \wedge p \in owner\}$
    \If{$OwnerBanks \neq \emptyset$}
        \State $b \gets \textproc{ChooseOwnerBank}(OwnerBanks,ReadPages)$
        \Comment{prefer fewer new read copies}
    \ElsIf{$\textproc{WriteQuotaFull}(current,WritePages,SPB)$}
        \State $b \gets \textproc{NextBank}(WritePages,ReadPages,SPB)$
    \Else
        \State $b \gets current$
    \EndIf
    \State $\textproc{PlacePages}(Pages,b)$
    \State $\textproc{AssignNewOwners}(WritePages,b)$
    \State $\textproc{MapIteration}(i,b)$; $current \gets b$
\EndFor
\end{algorithmic}
\end{algorithm}

\subsection{High-ILP Schedules}
\label{sec:schedule}

As discussed in Section \ref{sec:mothardware}, most NDP designs exhibit performance improvements due to the large amount of parallelism and bandwidth they have access to. However, we take it a step forward to improve the execution of each strand on the NDCores.
    
By observing loop execution, the SAE can analyze how the \UOPS are issued in an out-of-order (OoO) core, thus helping it create a high-performance ordering of \UOPS for the in-order NDCore. Our preliminary analysis observed that in-order NDCores offered ILP much lower than the out-of-order host cores. Therefore, we rely on SAE introspection to re-order \UOPS in the strand to achieve high ILP during its in-order execution.

A key observation is that when a sequence of x86 instructions executes on an in-order core, the ILP is extremely low because each x86 instruction is broken into multiple dependent \UOPS. The code inherently contains many back-to-back dependencies, which induce several stalls for the dependent \UOP and all subsequent \UOPS~. {\em The compiler cannot remove these stalls because the compiler works at the granularity of instructions, not \UOPS.} Efficient \UOP scheduling must, therefore, be performed by the hardware, and this is a unique opportunity where the hardware is creating \UOP strands and has the benefit of first observing how \UOPS were issued on an out-of-order core.

As the loop executes on the OoO core \circled{3} \circled{4}, the SAE tracks the order in which each \UOP is issued for $I$ iterations. The SAE then goes through this sequence to produce a \UOP listing for those iterations, which captures how the out-of-order core issued the \UOPS of the loop. This new \UOP schedule helps separate dependent \UOPS~, enabling much higher ILP through \UOP-level-parallelism when executing on the ND-Core. A higher value of $I$ produces a schedule using multiple iterations interleaved together for potentially higher performance at the cost of increasing the template size.

The above technique results in a strand with more ILP than is typically observed by in-order cores. Therefore, techniques like higher issue width, which have traditionally not been effective for in-order cores, can be more effective for NDCores. The impact of the above introspective schedule can, therefore, be amplified by combining it with an NDCore design that offers a higher issue width.

The generated \UOP schedule preserves register dependences. Memory operations are reordered only when MAT descriptors validate non-overlap for the accepted loop version; ambiguous aliases keep program order or cause the candidate to execute on the host.

\subsection{Supporting Coherence Between Offloads}
\label{sec:coherence}

As mentioned in Section~\ref{sec:looporchestration}, NDS benefits when the same loop is invoked multiple times and its page placement can be reused. To preserve correctness across those invocations, NDS uses a phase-based ownership model for pages managed by the LOU. When a matching Loop-Start \UOP commits, the host pauses at the loop boundary and the LOU acquires ownership of the pages that may be accessed by the offloaded strands.

Before the offload begins, cache lines from \NDR, \NDW, and \NDD pages are flushed or invalidated from the host caches, similar to selective flushing mechanisms~\cite{li2022fase}. This makes dirty host updates visible to memory before the LOU performs its mapping pass. During this pass, the LOU also refreshes \NDD pages so replicated read-only copies observe the latest host-visible values. Because the mapper records the banks that hold each replicated page, the LOU can bound this refresh work to replicated pages dirtied by intervening host execution rather than treating every read-only copy as stale. These steps give each NDCore a consistent starting view of the pages assigned to its strand.

During NDCore execution, the host does not access the acquired pages because the corresponding loop body is executing near memory. The page states from Section~\ref{sec:looporchestration} define the allowed access pattern: \NDR and \NDD pages are read-only for a strand, while each \NDW page has a single NDCore writer. When all strands complete, ownership returns to the host; any host cache lines for those pages must be refetched or revalidated before later host accesses. This explicit ordering between host memory operations and PIM execution follows the broader observation that PIM operations need a defined consistency contract with host loads and stores~\cite{perach2023consistency}.

Each accepted offload is blocking and runs to completion: the host waits at the loop boundary. The modeled design assumes stable physical mappings and no concurrent host access to acquired pages. Mid-offload interrupts, context switches, post-launch recovery, and arbitrary architectural live-outs are outside this model. Host fallback occurs only before launch.

\begin{table*}[tb]
\begin{tabular}{|l|l|l|l|}
\hline
\textbf{Requirement}                                                           & \textbf{Component}                                                              & \textbf{Verification Mechanism}                                                                                                                                                                                                                                                                                                                                                & \textbf{Scope and Limitations}                                                                                                                                                                                                                                                                           \\ \hline
\begin{tabular}[c]{@{}l@{}}Loop \\ Identification\end{tabular}                 & \begin{tabular}[c]{@{}l@{}}Loop \\ Detector \\ FSM\end{tabular}                 & \begin{tabular}[c]{@{}l@{}}Monitors dispatched branches and jumps to \\ detect two loop patterns (Fig 1). Once a \\ branch is identified \circled{2}, the FSM confirms \\ a loop candidate after 2 commits \circled{3}\circled{4}.\end{tabular}                                                                                                                                                     & \begin{tabular}[c]{@{}l@{}}Scope: Loops matching two patterns.\\ Excludes: Loops that commit $\le 2$\\ or have other complex control flow.\end{tabular}                                                                                                                                     \\ \hline
\begin{tabular}[c]{@{}l@{}}Sufficient \\ ND-Core \\ Resources\end{tabular}     & \begin{tabular}[c]{@{}l@{}}Dispatch \\ Buffer \& \\ LAE\end{tabular}            & \begin{tabular}[c]{@{}l@{}}During the second iteration \circled{3}\circled{4}, the LAE \\ collects metadata for each uop in the dispatch \\ buffer. An overflow of this buffer causes the \\ SAE to drop the candidate. During this \\ collection, the LAE also explicitly verifies that \\ 1) unique registers \textless ND-Core limit and 2) all \\ uops are on the supported list.\end{tabular} & \begin{tabular}[c]{@{}l@{}}Excludes: Loops that are too large, \\ have high register pressure, or\\  contain unsupported instructions.\\ (e.g., internal branches/if-else, \\ system calls).\end{tabular}                                                                                                   \\ \hline
\begin{tabular}[c]{@{}l@{}}Deterministic \\ Memory \\ Accesses\end{tabular}    & \begin{tabular}[c]{@{}l@{}}Memory \\ Access \\ Table \& \\ LAE\end{tabular}     & \begin{tabular}[c]{@{}l@{}}During the second iteration \circled{3}, the MAT \\ records addresses for all memory ops. It then \\ \circled{4} calculates the strides between consecutive \\ accesses and fails if the stride is not constant.\end{tabular}                                                                                                                                       & \begin{tabular}[c]{@{}l@{}}Scope: Constant-strided memory \\ accesses.\\ Excludes: Pointer-chasing and indexed \\ array accesses.\end{tabular}                                                                                                                                                              \\ \hline
\begin{tabular}[c]{@{}l@{}}Deterministic \\ Iteration \\ Space\end{tabular}    & \begin{tabular}[c]{@{}l@{}}Loop \\ Analysis \\ Engine\\ (LAE)\end{tabular}      & \begin{tabular}[c]{@{}l@{}}Performs backward dataflow analysis from \\ the loop's branch. It verifies the loop variable \\ is updated by a constant value and that the \\ loop limit is loop-invariant. \circled{3}\end{tabular}                                                                                                                                                           & \begin{tabular}[c]{@{}l@{}}Scope: Loop with simple inductions.\\ Excludes: Exit conditions are complex, \\ data-dependent, or modified within \\ the loop.\end{tabular}                                                                                                                            \\ \hline
\begin{tabular}[c]{@{}l@{}}Manage \\ Loop-Carried\\ Dependencies\end{tabular}  & \begin{tabular}[c]{@{}l@{}}Loop \\ Orchestration \\ Unit \\ (LOU)\end{tabular}  & \begin{tabular}[c]{@{}l@{}}As the LOU maps iterations to execution \\ banks \circled{5}, it tracks page-level write \\ ownership, co-locates dependent work at \\ owner banks, and rejects unsafe state \\ transitions before offload.\end{tabular}                                                                                                                                   & \begin{tabular}[c]{@{}l@{}}Scope: Handles dependencies \\ expressible through page ownership.\\ Excludes: Dependences not captured \\ by the page-level model.\end{tabular}                                                                             \\ \hline
\begin{tabular}[c]{@{}l@{}}Maintain Host\\ Coherence\end{tabular}              & \begin{tabular}[c]{@{}l@{}}LOU @\\ Every\\ Invocation\end{tabular}              & \begin{tabular}[c]{@{}l@{}}At the loop boundary, the LOU acquires \\ the managed pages, flushes or invalidates \\ host cache lines, refreshes replicated pages, \\ and releases ownership after completion.\end{tabular}                                                                                                                                                   & \begin{tabular}[c]{@{}l@{}}Scope: Phase-based ownership for \\ pages used by the offloaded loop.\end{tabular}                                                                                                                                                                                              \\ \hline
\begin{tabular}[c]{@{}l@{}}Manage \\ Multiple Loop \\ Versions\end{tabular} & \begin{tabular}[c]{@{}l@{}}LST @\\ Dispatch /\\ LHT+VVS @\\ Commit\end{tabular} & \begin{tabular}[c]{@{}l@{}}Every uop is looked up in the Loop Start Table\\ (LST) on dispatch. A hit triggers a single\\ lookup in the Loop Header Table (LHT) at \\ commit. This LHT lookup finds the matching\\ version in the Version Value Store (VVS) to\\ start the offload\end{tabular}                                                                                  & \begin{tabular}[c]{@{}l@{}}Scope: Validates version matching.\\ Excludes: Number of entries in the \\ LST puts an upper limit on the \\ number of offloads at a time. Loops \\ live-in values exceed the \\  VVS entry capacity.\end{tabular} \\ \hline
\end{tabular}
\caption{Summary of checks made by the NDS Framework to validate offload eligibility}
\label{tab:checks}
\end{table*}

\subsection{Supporting Nested Loops}

So far, our discussion has only considered single-level loops. However, it is crucial to consider multi-level nested loops when applicable. A significant portion of the offload cost for subsequent outer iterations can be avoided by performing a single offload for the entire loop nest. Doing so eliminates potential redundancy in back-to-back offloads by improving the loop orchestration.

To support nested loops, we require the following additions to the framework:

\subsubsection*{Strand Detection and Analysis}  When a potential candidate loop during the profiling phase \circled{3} \circled{4} encounters another loop, it is immediately tagged as an outer loop. The SAE parameterizes the loop variable and stores only the dispatched \UOPS associated with the outer loop. Moreover, outer loops which involve stores made to live-in registers/addresses not associated with a \textit{LV} are rejected as candidates for nesting. This is done to avoid the added complexity at the NDCore to support the nesting. The live-in registers and addresses are tracked for the loop nest as a whole and are encoded in the strand when they are sent to the NDCore.

\subsubsection*{Loop Orchestration and Data Marshaling} 
\begin{figure}
  \begin{minipage}{0.4\linewidth}
    \includegraphics[width=\linewidth]{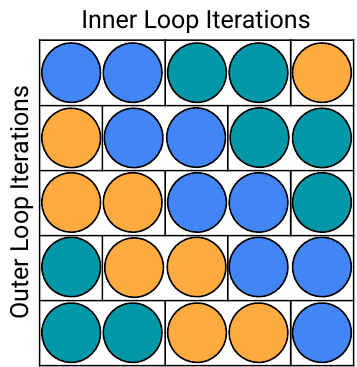}
  \end{minipage}\hfill
  \begin{minipage}[c]{0.59\linewidth}
    \caption{
       Example of a 2-level nested loop (iterations as circles) organized into strands (grouped in boxes) and partitioned across 4 banks (different colors). Each bank parallely executes the strands assigned to them. Maximum iterations executed in one bank is 9 (Blue).
    } \label{fig:iterationspace}
    \Description{Two-dimensional nested-loop iteration space partitioned into strand groups and assigned across four memory banks.}
  \end{minipage}
\end{figure}

Each nest of the loop tracks the stride for memory access, so when the offloading unit makes offloading decisions, it can co-locate all pages and iterations for all levels of the nesting at the same time using the same orchestrating logic. This partitioning can be visualized using Figure \ref{fig:iterationspace}, which shows an example of how an iteration space can be divided across NDCores.


\color{black}

\section{Methodology}
\label{sec:method}

\subsection{Implementation Details}
\label{sec:impldetails}

\begin{table}[tb]
\small
\begin{tabular}{|ll|ll|l|}
\hline
\multicolumn{1}{|l|}{Structure}                                                                          & Sub Structure               & \multicolumn{1}{l|}{Size} & Entries & Total Size                  \\ \hline
\multicolumn{2}{|c|}{Loop Start Table}                                                                                                  & \multicolumn{1}{l|}{73 b} & 32      & 0.3 KB                      \\ \hline
\multicolumn{1}{|l|}{\multirow{3}{*}{\begin{tabular}[c]{@{}l@{}}Strand\\ Analysis\\ Engine\end{tabular}}} & Loop Detection        & \multicolumn{1}{l|}{99 b} & 3       & \multirow{3}{*}{$\sim$4 KB} \\ \cline{2-4}
\multicolumn{1}{|l|}{}                                                                                    & Memory Access         & \multicolumn{1}{l|}{81 b} & 64      &                             \\ \cline{2-4}
\multicolumn{1}{|l|}{}                                                                                    & \UOP Buffer & \multicolumn{1}{l|}{64 b} & 512     &                             \\ \hline
\multicolumn{1}{|l|}{\multirow{5}{*}{\begin{tabular}[c]{@{}l@{}}Max\\ Template\\ Size\end{tabular}}}      & Loop Body                   & \multicolumn{1}{l|}{64 b} & 512     & \multirow{5}{*}{$\sim$5 KB} \\ \cline{2-4}
\multicolumn{1}{|l|}{}                                                                                    & Issue Order                 & \multicolumn{1}{l|}{3 b}  & 512     &                             \\ \cline{2-4}
\multicolumn{1}{|l|}{}                                                                                    & Live-in IDs                 & \multicolumn{1}{l|}{16 b} & 1       &                             \\ \cline{2-4}
\multicolumn{1}{|l|}{}                                                                                    & Live-in Values              & \multicolumn{1}{l|}{64 b} & 12      &                             \\ \cline{2-4}
\multicolumn{1}{|l|}{}                                                                                    & Memory Accesses             & \multicolumn{1}{l|}{64 b} & 64      &                             \\ \hline
\end{tabular}
\caption{Size of storage structures used on the host core. Stored addresses use 48-bit fields; the 81-bit memory-access entry is a 48-bit address, 1-bit access type, and 32-bit stride.}
\label{tab:size}
\end{table}
The NDS framework is designed with lightweight logic for its Strand Detection, Analysis, and Orchestration components. Much of the hardware overhead in the host is because of metadata storage. 
Table~\ref{tab:size} summarizes the sizes of the different metadata structures. 
We synthesize the host-side control RTL using Synopsys Design Compiler~\cite{synopsysDesignCompiler} at a 12-nm technology node and model larger metadata arrays using GF12 register-file and SRAM macros; the current host frontend occupies 0.037~mm$^2$, including 0.016~mm$^2$ of modeled metadata storage.
The area estimate therefore combines synthesized control logic with technology-specific macro models for the storage structures.

\subsubsection*{Strand Analysis Engine} The SAE consists of the Loop Detection FSM and Loop Analysis Engine (Figure~\ref{fig:host_hardware}) and is present on every core. It tracks only one inner loop and its corresponding nested structures at a time, flushing all associated data upon completion of strand analysis. As a result, the Loop Detection Table, Memory Access Table, and \UOP buffer are sized by the maximum nesting depth, number of distinct memory streams, and maximum loop body that can be captured. We provision 512 \UOP entries to cover a logical maximum of 500 loop-body \UOPS with power-of-two rounding. The template also stores the observed issue order separately from the loop body because this order is what enables Smart Schedule execution on the in-order NDCore. Live-in identity metadata is likewise separate from live-in values: the former records which architectural values must be checked or captured, while the latter stores the actual values used to instantiate a strand. Considering these factors, we estimate that the SAE introduces a minor overhead compared to existing Loop Stream Detectors. 

\subsubsection*{Loop Orchestration Unit} The LOU performs the loop orchestration and data marshaling required by every offload and is present on each memory controller. It stores all the generated templates and their associated metadata in a scratchpad that can be handled similarly to the trace caches in \cite{padmanabha2017mirage}. The size of this scratchpad puts an upper limit on the number of different templates that can be offloaded during the program execution.

\begin{table}[ht]
\centering
\begin{tabular}{|c|c|l|}
\hline
Configuration &
  Type &
  Description \\ \hline
\multirow{2}{*}{\begin{tabular}[c]{@{}l@{}}Location of\\ Processor\end{tabular}} &
  Host &
  \begin{tabular}[c]{@{}l@{}}8 Cores\\ L1: 32 KB, 8-way \\ L2: 256 KB, 8-way \\ L3: 8 MB, 12-way\\ 45 ns DRAM access latency\end{tabular} \\ \cline{2-3} 
 &
  NDP &
  \begin{tabular}[c]{@{}l@{}}1 Core per Bank (256 Banks)\\ L1: 64 KB, 8-way \\ 30 ns DRAM access latency\end{tabular} \\ \hline
\multirow{3}{*}{\begin{tabular}[c]{@{}l@{}}Mode of\\ Operation\end{tabular}} &
  OoO &
  Out of Order \\ \cline{2-3} 
 &
  InO &
  In Order \\ \cline{2-3} 
 &
  SS+ &
  Smart Schedule Plus \\ \hline
\multirow{2}{*}{\begin{tabular}[c]{@{}l@{}}Core\\ Resources\end{tabular}} &
  Beefy &
  \begin{tabular}[c]{@{}l@{}} 3.6 GHz Cores\\ 4 wide, 224-entry\\ L2 GHB prefetcher\end{tabular} \\ \cline{2-3} 
 &
  Wimpy &
  \begin{tabular}[c]{@{}l@{}} 350 MHz Cores\\ 2 wide, 128-entry \end{tabular} \\ \cline{2-3} 
 \hline
\end{tabular}
\caption{Configuration parameters for the design space.}
\label{tab:parameters}
\vspace{-10pt}
\end{table}

\subsection{Modeling the Hardware}
\label{sec:model}

We evaluate the proposed NDS framework against a Coffee-Lake baseline system, as described in Table \ref{tab:parameters}. We consider many configurations that are described next. 

\textit{Location of Processor} refers to whether computation is on host or memory. The host processor is the baseline and has access to three levels of cache and the normal DRAM access latency. An NDP processor only sees a single cache level with reduced DRAM access latency. When the host accesses memory, the DRAM latency is about 45~ns (not including wait times for the memory channel to be free). 
The column access for an NDCore is significantly faster since it does not involve data movement to the pins and over the channel. Hence, the DRAM latency for the NDCore is assumed to be 30~ns. 

\textit{Mode of Operation} of the processor can be either (1) In Order, where the executing instructions (and by extension their corresponding \UOPS) are processed in the exact order they appear in the program, (2) Out of Order, where execution allows the executing \UOPS to be executed out of sequence based on dependencies and resource availability, and (3) Smart Schedule, where the framework learns a schedule for efficient execution from the OoO host and uses that instead of the program order to issue \UOPS.

\textit{Core Resources} can either be beefy, which is based on the Coffee-Lake Client \cite{intelcl,chipscl} system, or wimpy, which models the limited resources of a processor near memory banks.

\subsection{Workloads}
As our work targets general-purpose applications that run on the CPU, we take workloads from a wide range of benchmark suites. We consider programs from the STREAM \cite{mccalpin1995stream} (Copy, Scale, Add, Triad), Splash-3 \cite{sakalis2016splash} (Ocean-NCP \& CP, LU-CB \& NCB), and Polybench \cite{karimov2019polybench} (GEMVER, Jacobi-1D and 2D) benchmark suites. For each program, we use the large input dataset size, summarized in Table~\ref{tab:workloads}. We focus on the main kernels in these workloads. We also perform this analysis for the Multi-Head Attention (MHA) kernel present in a Large Language Model (LLM) llama2.c \cite{llama2c} running the TinyLlama-1.1B model \cite{tinyLlama}; we measure the time taken to produce the 2048th token. 

\begin{table}[ht]
\centering
\begin{tabular}{|l|lll|}
\hline
Suite & \multicolumn{1}{l|}{Workload} & \multicolumn{1}{l|}{Kernel} & Input \\ \hline
\multirow{4}{*}{STREAM} & STREAM & Copy & 10000000 \\
 & STREAM & Scale & 10000000 \\
 & STREAM & Add & 10000000 \\
 & STREAM & Triad & 10000000 \\ \hline
\multirow{4}{*}{Splash-3} & Ocean-NCP & Jacob & n1026 \\
 & Ocean-CP & Jacob & n1026 \\
 & LU-CB & Bmod & n1024 b512 \\
 & LU-NCB & Bmod & n1024 b512 \\ \hline
\multirow{3}{*}{Polybench} & \multicolumn{2}{c}{Gemver} & LARGE \\
& \multicolumn{2}{c}{Jacobi-1D} & LARGE \\
& \multicolumn{2}{c}{Jacobi-2D} & LARGE \\ \hline
llama2.c & \multicolumn{2}{c}{Multi-Head Attention}  & TinyLlama-1.1B \\ \hline
\end{tabular}
\caption{Workloads used in the study with their input size}
\label{tab:workloads}
\end{table}

\subsection{Simulation Setup}
We use the Sniper simulator \cite{carlson2014evaluation} to measure our execution metrics. We study both the full kernel runs and the inner loops in them in isolation. Table~\ref{tab:parameters} summarizes many of our simulator parameters. Using the OoO runs, we generate \UOP traces that are fed to the SAE and LOU. When using the in-order core with a smart schedule, the modified \UOP schedule is fed to Sniper.


We estimate the page movement cost using the reported CPU-DPU bandwidth from UPMEM~\cite{gomez2021benchmarking}, which is a conservative estimate for modeling these costs. We estimate the worst-case cost of a flush as the ratio of total cache size to effective DRAM write bandwidth, $10.25 MB/(41.6 GBs^{-1}\times80\%)=308\mu s$, and use that for every offload. We also instrument the workload with timestamps to calculate the cost and impact of the flushes on the rest of the kernel execution. The CPU baselines are the out-of-order host and 8-core host-parallel variants where available, using the same input sizes, measured kernel regions, and timestamp method as NDS.

\section{Results}
\label{results}

\subsection{Strand Detection and Analysis}

\begin{figure}[ht]
    \centering
    \includegraphics[width=\linewidth]{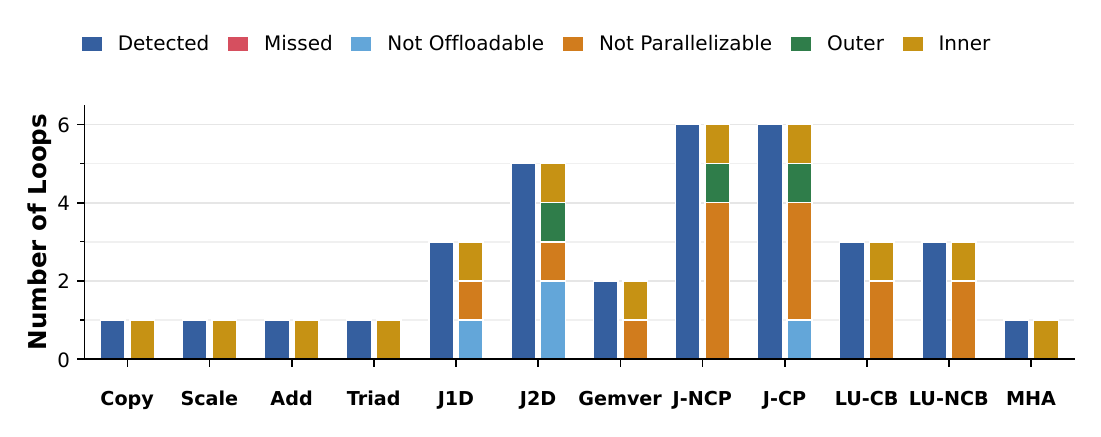}
    \caption{SAE analysis of the different loops. The first bar breaks loops as detected or missed. The second bar breaks loops as Not-offloadable, Not-parallelizable, and as Outer/Inner (if Offloadable and Parallelizable).}
    \Description{Stacked bar chart showing detected and missed loops, then classifying detected loops by offloadability, parallelizability, and inner or outer loop status.}
    \label{fig:res_charac}
\end{figure}

Figure~\ref{fig:res_charac} shows the effectiveness of the SAE's loop analysis. Across the instrumented kernel regions, the SAE detects all loops matching the branch patterns in Figure~\ref{fig:loop_examples}. Of 33 detected loops, 12 inner loops and 3 outer loops satisfy the offloadability and parallelizability checks, while 14 are rejected as not parallelizable and 4 as not offloadable.

Every workload exposes at least one inner-loop candidate. Rejections occur when the SAE cannot construct a safe page mapping, dependences conflict with a previous offload, the loop body exceeds template constraints, or the expected work is too small to amortize offload overheads.

\begin{figure*}[!t]
    \centering
    \includegraphics[width=0.9\linewidth]{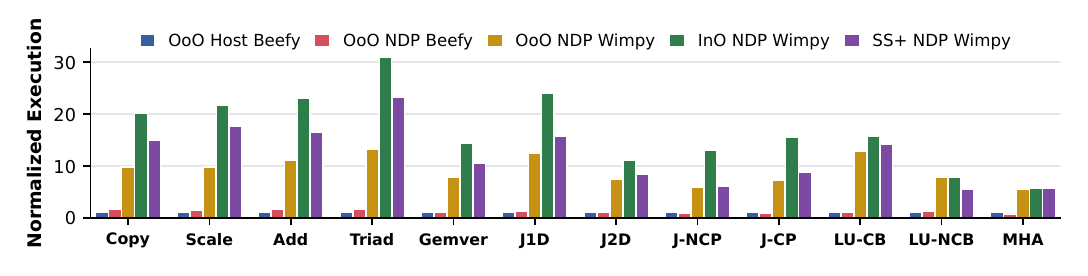}
    \caption{Effect of processor type on the execution time of the innermost candidates, normalized to OoO Host Beefy.}
    \Description{Bar chart comparing normalized execution time for host and near-data execution under out-of-order, in-order, and smart-scheduled in-order processor configurations.}
    \label{fig:res_loctype}
\end{figure*}

\subsection{Effect of Smart Schedules and Issue Width}

We first analyze Smart Schedule independently of full NDS parallelization because the near-data cores are deliberately simple. Smart Schedule uses OoO host issue order as a guide for later in-order NDCore execution, while \textit{SS+} also checks the oldest \UOP in program order to avoid stalls when the learned schedule cannot issue immediately. In \textit{STREAM-ADD}, plain InO execution reaches only 0.24$\times$ of OoO performance; Smart Schedule improves this to 0.34--0.46$\times$.

SS+ consistently improves over the simpler policy, extra profiled iterations provide only small gains, and the starting iteration has little visible effect. We therefore use compact one-iteration schedules. Issue width shows a similar diminishing return: width 2 gives a 1.45$\times$ average benefit over width 1, while width 4 gives 1.56$\times$. The remaining evaluations use SS+ with width 2.

\subsection{Location and Type of Execution}

Figure~\ref{fig:res_loctype} isolates the cost of executing the selected innermost candidates on different processor locations and core types before exploiting bank-level parallelism. Moving an out-of-order beefy core from the host to the NDP location is not enough to produce large gains by itself: the NDP beefy configuration has a 1.16$\times$ geomean runtime relative to the host beefy baseline. This result reflects the tradeoff between lower memory access latency near DRAM and the shallower memory hierarchy and lower controller bandwidth available to the near-data core.

The more important effect is the reduced capability of practical NDCores. Replacing the beefy core with a wimpy out-of-order NDCore increases runtime to 8.84$\times$ geomean over the host baseline, and executing the same strands on a plain in-order wimpy NDCore increases runtime to 15.31$\times$ geomean. SS+ recovers part of this loss by using the issue order observed on the out-of-order host to improve in-order issue on the NDCore: SS+ reduces the geomean runtime to 10.99$\times$, a 1.39$\times$ geomean speedup over plain in-order execution. Thus, learned schedules make each simple NDCore more effective, while the bank-level parallelism evaluated next provides the additional scaling needed to outperform host execution.

\subsection{Effect of Parallelism}

\begin{figure*}[ht]
    \centering
    \includegraphics[width=0.95\linewidth]{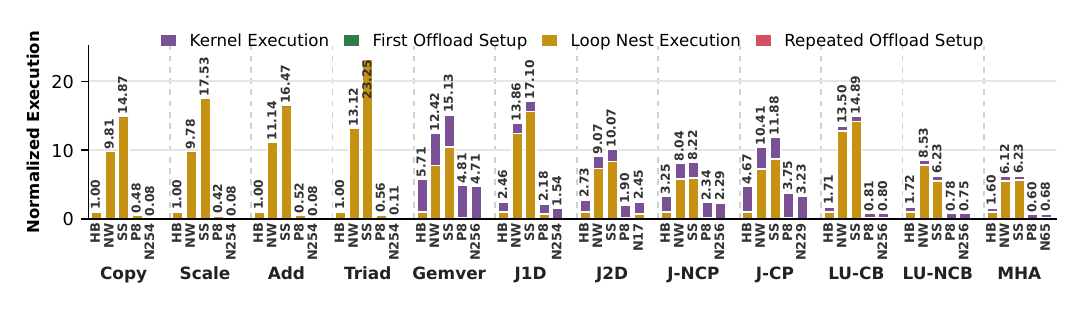}
    \vspace*{-10pt}
    \caption{Effect of parallelism on normalized execution time. The first three bars are non-parallel design points from Figure~\ref{fig:res_loctype}. Px indicates the loop nest was run in parallel across x host cores. Ny indicates that it was run in parallel across y NDCores.}
    \Description{Bar chart showing normalized execution time as host-core and near-data-core parallelism increase.}
    \label{fig:mainchart}
\end{figure*}

Figure~\ref{fig:mainchart} demonstrates the costs and benefits of parallelizing accepted loop nests across multiple NDCores. Relative to the OoO host baseline for each workload, NDS improves performance by 3.14$\times$ geomean and 5.03$\times$ arithmetic mean, with speedups ranging from 1.11$\times$ to 13.23$\times$. Compared with 8-core host-parallel variants, NDS improves performance by 1.82$\times$ geomean and 2.60$\times$ arithmetic mean, and outperforms the host-parallel version on 10 of the 12 workloads.

The largest gains occur when the selected loop accounts for most of the measured execution. The \textit{STREAM} kernels spend nearly all of their measured time in the offloaded loop, so NDS achieves 13.23$\times$, 11.88$\times$, 12.43$\times$, and 9.23$\times$ speedups for Copy, Scale, Add, and Triad, respectively. Workloads with substantial non-offloaded remainder are bounded by Amdahl's law. For example, the selected loop accounts for 17.5\% of the measured \textit{Gemver} execution, 21.4\% of \textit{Jacob CP}, 30.8\% of \textit{Jacob NCP}, and 36.7\% of \textit{Jacobi 2D}. NDS still improves whole-kernel execution for these workloads, but the remaining host execution limits the end-to-end gain.

The Loop Orchestration Unit is responsible for splitting the target loop across NDCores while enforcing coherence and minimizing page duplication. As a result, the effective parallelism varies by benchmark and can be lower than the 256 NDCores available in the modeled system. Across workloads, NDS uses 202 effective NDCores on average, but the range is wide: \textit{Jacobi 2D} uses only 11.7 effective NDCores, MHA uses 64, \textit{Jacob NCP} uses 171.6, \textit{Jacob CP} uses 228, and the \textit{STREAM} kernels use about 254. This variation comes from the original page allocation; because NDS is transparent, it cannot remap the application data structures to align them perfectly with NDP banks before the first offload.

The two workloads where NDS trails the 8-core host-parallel baseline illustrate the limits of the approach. \textit{Jacobi 2D} exposes only 11.7 effective NDCores because its page mapping limits independent near-data partitions. MHA improves over the single-threaded host baseline but trails the host-parallel baseline because the selected loop uses only 64 effective NDCores and the host-parallel implementation is already effective.

\subsection{Amortizing Offload Costs}

\begin{figure}
    \centering
    \includegraphics[width=\linewidth]{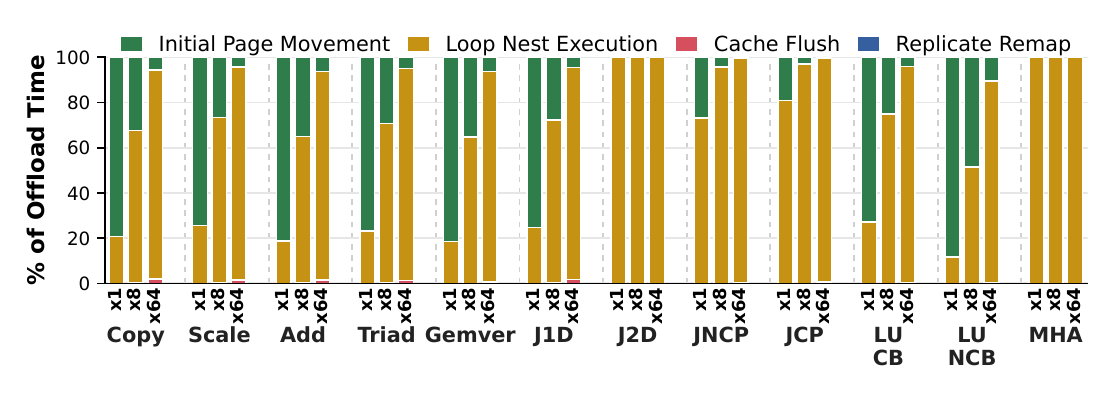}
    \vspace*{-6pt}
    \caption{$\%$ of time taken by each component of an NDCore offload as the number of offloads increase.}
    \Description{Stacked bar chart showing how NDCore offload time is divided among components as offload count increases.}
    \label{fig:spread}
\end{figure}

To offload strands, NDS pays a one-time page movement cost during the data marshaling step of the first offload, then pays recurring costs to maintain coherence using flushes and page movements for later offloads. These costs depend on the original data mapping and on the number of NDCores used for parallel execution. Figure~\ref{fig:spread} breaks down how those costs change as the same strand is invoked repeatedly.

Averaged across workloads, initial page movement accounts for 56.3\% of the one-call offload time, while useful loop execution accounts for 43.7\%. Using the component data in Figure~\ref{fig:spread}, the one-time page movement cost falls below useful loop execution within 8 repeated invocations for every workload, and within 4 invocations for 9 of 12 workloads. On average, the initial movement share drops to 22.3\% at 8 invocations and 3.9\% at 64 invocations. At that point, useful loop execution accounts for 95.2\% of offload time, with cache flushes contributing less than 1\% on average. We observe no replicated-page remapping cost because the host does not modify replicated pages between offloads.

These results explain why NDS is most effective for repeated regular loops. The hardware can identify and offload a loop without programmer involvement, but the offload becomes most attractive when the template and page movement costs can be reused across many dynamic invocations. This is the common case for streaming kernels and iterative stencil-like computations, and it is also the regime where better OS or runtime page placement support could further improve transparent NDP performance without changing the binary.

\section{Related Work and Future Directions}
\label{sec:related}

\noindent\textbf{Support for strand detection.}
Proposals like \cite{devic2022topim, wei2022pimprof,oliveira2021damov,kim2017toward,ghiasi2022alp,maity2023data,ahmed2019compiler} aim to identify code regions suitable for near-data execution. However, they rely on an offline profiling step (during compile-time or run-time) that either demands extensive analytics and/or assumes access to application source code for compilation, a requirement that can't be consistently guaranteed for users. Our proposal uses a lightweight in-flight mechanism to detect and offload strands to the NDCores, which is possible due to NDS's focus on loops. Works from the past decade that focus on improving the performance of loops (~\cite{moseley2006loopprof, clark2008veal} for loop identification and/or software-hardware co-design), along with other compiler based techniques \cite{maity2023data,yan2021copim,lee2001automatically,jiang20243,kandemir2021compiler,lockerman2020livia,schwedock2022tako} require modifications to the software stack and/or have the potential to introduce large amounts of overheads to the actual execution of the program. It is important to note that our framework has the ability to make use of these proposals by coupling the SAE with the infrastructure they propose to leverage a wider and richer range of strands. We are further able to target workloads not amenable to offline profiling.

\noindent\textbf{Support for NDP.}
Most other NDP proposals introduce special hardware structures near cache/memory (or suggest changes to the hardware itself) to improve the efficiency of a few operations \cite{shafiee2016isaac,seshadri2017ambit,lockerman2020livia,kim2021aquabolt, wang201928computesram}, design specifically for specific workload types \cite{wang2022stream, kwon2023skhynix, gu2020ipim}, or rely heavily on programmer involvement \cite{devaux2019upmem} to improve performance. Most of these proposals also suggest ISA extensions to better utilize the underlying NDP hardware. Instead, our work proposes a seamless mechanism without any changes to the ISA to detect Strands and offload them to the memory hierarchy while operating at a coarser granularity of loops. Moreover, our framework can be used to simplify programming and data marshaling for proposals like \cite{kim2021aquabolt} by adding support for vectorization. Our framework can also use ideas from other proposals like \cite{boroumand2019conda,hashemi2016emcyale,fujiki2023mvc,wang2020figaro,wang2023affinity,lockerman2020livia,pugsley2014comparing,pugsley2014ndc,rai2021design,alian2018application,lee2024pim, noh2024pid} and upcoming paradigms like CXL \cite{cxl} to overcome its shortcomings.
Note that near-memory-processing offers high bandwidth and parallelism, whereas competing approaches like GPUs offer less memory bandwidth, and approaches like prefetching only address latency, but not bandwidth.

\noindent\textbf{Support for efficient in-order execution.}
Most NDP frameworks see benefits in performance due to the lower memory latency, high bandwidth and/or parallelism they can provide, in spite of the poor processing capabilities of the NDCores themselves. Moreover, as discussed in Section \ref{sec:mothardware}, there is a large scope for increasing support to improve the in-order execution of these NDCores. Our framework uses an idea similar to that of trace caches and proposals similar to it \cite{sharifian2016chainsaw, fallin2014heterogeneous, brandalero2016potential, brandalero2017mechanism, nair1997exploiting}, and more specifically techniques \cite{padmanabha2015dynamos,padmanabha2017mirage} that learn execution patterns on the host to improve offloaded performance on different compute units. Unlike previous approaches, our framework is unique because of the focus on loops and their inherent regularity. This allows us to condense the size of traces needed to infer extremely large execution patterns into a comparable area overhead. Also note that our hardware based approach enables us to work with micro-ops which is not possible for most prior approaches.

\color{black}

\section{Conclusions}
\label{sec:concl}

This paper introduces Near Data Strands (NDS), an architecture that automatically and transparently identifies, validates, parallelizes, and offloads parts of the program during run time for near-memory execution. We discuss the hardware structures required to support the framework and also describe a method to improve near-data execution using schedules learned from the out-of-order execution. We thus show that NDS can identify and analyze simple loops during runtime and extract high NDP performance with a combination of smart instruction scheduling and data marshaling aware parallelization. While highly effective for dense linear algebra and strided array processing, the approach explicitly excludes loops with complex, data-dependent control flow (e.g., pointer chasing) or non-constant access strides that prevent safe parallelization (e.g., Sparse Matrix Vector Multiplication). The modeled design further excludes mid-offload interrupts and context switches, post-launch recovery, and arbitrary architectural live-outs.

\begin{acks}
We thank the reviewers. This work was supported in part by \grantsponsor{GS100000001}{NSF}{https://www.nsf.gov/} grant \grantnum{GS100000001}{CCF-2217154} and \grantsponsor{intel}{Intel}{https://www.intel.com/}.
\end{acks}

\bibliographystyle{ACM-Reference-Format}
\bibliography{refs}

\end{document}